\documentclass[preprint]{elsarticle}

\usepackage{hyperref}
\usepackage{subfigure}
\usepackage{amsmath}
\usepackage{lineno}
\hypersetup{colorlinks = true,
            linkcolor = blue,
            anchorcolor = red,
            citecolor = blue,
            filecolor = red,
            urlcolor = red,
            pdfauthor = author}

\journal{Nucl. Instrum. Methods A}

\begin{document}

\begin{frontmatter}

\title{In-orbit Radiation Damage Characterization of SiPMs in the GRID-02 CubeSat Detector}

\author[THU_KeyLab,THU_DEP]{Xutao Zheng}
\author[THU_KeyLab,THU_DEP]{Huaizhong Gao}
\author[THU_KeyLab,THU_DEP]{Jiaxing Wen}

\author[THU_KeyLab,THU_DEP]{Ming Zeng\corref{CA}}
\cortext[CA]{Corresponding author}
\ead{zengming@tsinghua.edu.cn}

\author[THU_KeyLab,THU_DEP]{Xiaofan Pan}
\author[THU_DEP]{Dacheng Xu}
\author[THU_DEP]{Yihui Liu}
\author[THU_DP]{Yuchong Zhang}
\author[THU_DP]{Haowei Peng}
\author[THU_DEP]{Yuchen Jiang}
\author[THU_KeyLab,THU_DEP]{Xiangyun Long}
\author[THU_DEP]{Di'an Lu}
\author[THU_KeyLab,THU_DEP]{Dongxin Yang}
\author[THU_KeyLab,THU_DEP,THU_DA]{Hua Feng}
\author[THU_KeyLab,THU_DEP]{Zhi Zeng}
\author[THU_DA]{Jirong Cang}
\author[THU_KeyLab,THU_DEP]{Yang Tian}
\author[]{GRID Collaboration}

\address[THU_KeyLab]{Key Laboratory of Particle and Radiation Imaging (Tsinghua University), Ministry of Education, Beijing, 100084, People's Republic of China}
\address[THU_DEP]{Department of Engineering Physics, Tsinghua University, Beijing, 100084, People's Republic of China}
\address[THU_DP]{Department of Physics, Tsinghua University, Beijing, 100084, People's Republic of China}
\address[THU_DA]{Department of Astronomy, Tsinghua University, Beijing, 100084, People's Republic of China}

\begin{abstract}
    Recently, silicon photomultipliers (SiPMs) have been used in several space-borne missions, owing to their solid state, compact size, low operating voltage, and insensitivity to magnetic fields. 
    However, operating SiPMs in space results in radiation damage and degraded performance. 
    In-orbit quantitative studies on these effects are limited. 
    In this study, we present in-orbit SiPM characterization results obtained by the second detector of the Gamma-Ray Integrated Detectors (GRID-02), which was launched on 6 November 2020. 
    An increase in dark current of $\sim$100 $\mu$A/year per SiPM chip (model MicroFJ-60035-TSV) at 28.5 V and 5 $^{\circ}$C was observed. 
    Consequently, the overall noise level (sigma) of the GRID-02 detector increased by $\sim$7.5 keV/year. 
    The estimate of this increase is $\sim$40 $\mu$A/year per SiPM chip at -20 $^{\circ}$C, highlighting the positive effect of using a cooling system. 
\end{abstract}


\end{frontmatter}

\section{Introduction}

Silicon photomultipliers (SiPMs) have recently become popular photodetectors for scintillator readout. 
Thanks to their compact size, solid state, low operating voltage, and insensitivity to magnetic fields, they have become practical alternatives to traditional photomultiplier tubes (PMTs), especially for space-borne applications. 
As summarized in Table \ref{tab:missions}, several space missions equipped with SiPMs have been launched. 
SiPMs were used for scintillator readout \cite{Bencardino2005449, LI201663, SIRI-1, mitchell2019strontium, SIRI-2, ZHANG20198, GRBAlpha, CAMELOT, GARI} or ultraviolet (UV) observation \cite{CAPEL20182954}. 
However, it is known that SiPMs are susceptible to radiation damage in space environment, which leads to significant increase in dark count rate and degrades their noise performance. 
Few in-orbit radiation damage effect studies of SiPMs have been reported, although many experiments have been conducted at accelerator facilities \cite{LI201663, BARTLETT2020163957, MITCHELL2021164798, HIRADE2021164673, ULYANOV2020164203}. 
SIRI-1 reports one of the few in-orbit results showing a significant increase in the dark current of SiPM with respect to time and dose \cite{MITCHELL2021164798}. 

\begin{table}
    \begin{center}
        \setlength{\leftskip}{-4pt}
        \caption{Launched space missions equipped with SiPMs. }
        \begin{tabular}{cccc}
            \hline
            Project name        & Launch date   & Orbit                                     & Ref.                                  \\ 
            \hline
            Lazio-Sirad         & 2005.04.15    & 417 $\times$ 423 km 51.6$^{\circ}$ ISS    & \cite{Bencardino2005449}              \\ %
            HXMT                & 2017.06.15    & 533 $\times$ 543 km 43.0$^{\circ}$ LEO    & \cite{LI201663}                       \\ 
            GRID-01             & 2018.10.29    & 502 $\times$ 519 km 97.4$^{\circ}$ LEO    & \cite{Wen2019}                        \\ 
            SIRI-1              & 2018.12.03    & 567 $\times$ 589 km 97.7$^{\circ}$ SSO    & \cite{SIRI-1, mitchell2019strontium}  \\ 
            Mini-EUSO           & 2019.08.22    & 417 $\times$ 423 km 51.6$^{\circ}$ ISS    & \cite{CAPEL20182954}                  \\ %
            LabOSat             & 2020.01.15    & Unknown LEO                               & \cite{BARELLA2020164490}              \\ 
            GRID-02             & 2020.11.06    & 457 $\times$ 467 km 97.2$^{\circ}$ LEO    & \cite{Wen2019}                        \\ 
            GECAM               & 2020.12.09    & 586 $\times$ 604 km 29.0$^{\circ}$ LEO    & \cite{ZHANG20198}                     \\ 
            GRBAlpha            & 2021.03.22    & 534 $\times$ 563 km 97.5$^{\circ}$ LEO    & \cite{GRBAlpha}                       \\ 
            SIRI-2              & 2021.12.07    & 35784 $\times$ 35788 km 0.0$^{\circ}$ GEO & \cite{SIRI-2}                         \\ 
            GARI                & 2021.12.21    & 417 $\times$ 423 km 51.6$^{\circ}$ ISS    & \cite{GARI}                           \\ %
            GRID-03B \& GRID-04 & 2022.02.27    & 523 $\times$ 550 km 97.5$^{\circ}$ LEO    & \cite{Wen2019}                        \\ 
            \hline
        \end{tabular}
        \label{tab:missions}
    \end{center}
\end{table}

The Gamma-Ray Integrated Detectors (GRID) \cite{Wen2019} is a space mission concept with the purpose of monitoring gamma-ray bursts (GRBs) using a constellation of CubeSats equipped with a scintillator detector readout by SiPM. 
It is also specifically designed with SiPM performance study setups to conduct SiPM characterization experiments. 
The first GRID detector, GRID-01, was launched on 29 October 2018 and the subsequent one, GRID-02, was launched on 6 November 2020. 
Many other similar missions are under development, including BurstCube \cite{racusin2017burstcube}, HERMES \cite{FUSCHINO2019199}, EIRSAT-1 (GMOD) \cite{Murphy2021}, Glowbug \cite{Grove2019109}, CUBES \cite{Kushwah_2021}, and GALI \cite{GALI}, most of which will use SiPM to read out scintillators. 

In this article, we present the in-orbit SiPM characterization results obtained by GRID-02 during the first few months after launch. 
Based on the measurement setups of the detector, we analyzed the change in breakdown voltage and dark current due to radiation damage. 
An estimate of the dark current increase rate of the SiPM operating in low earth orbit (LEO) was given. 
The energy resolution and low energy threshold deterioration of GRID-02 as a result of dark count noise increase were also studied. 

\section{In-orbit characterization setup and methods}

In addition to the optimization of gamma-ray detection in GRID, several functions have been specifically designed for performance studies of SiPMs. 
The operating conditions of the SiPMs, including the current, bias voltage, and temperature, were monitored. 
In addition, a charge injection module was used to analyze the noise contributions. 
The details of the GRID instrument design can be found in \cite{Wen2021}. 
Herein, we briefly review the biasing and readout circuits, and focus on the characterization setup. 

\begin{figure}
    \centering
    \includegraphics[width=0.8\textwidth]{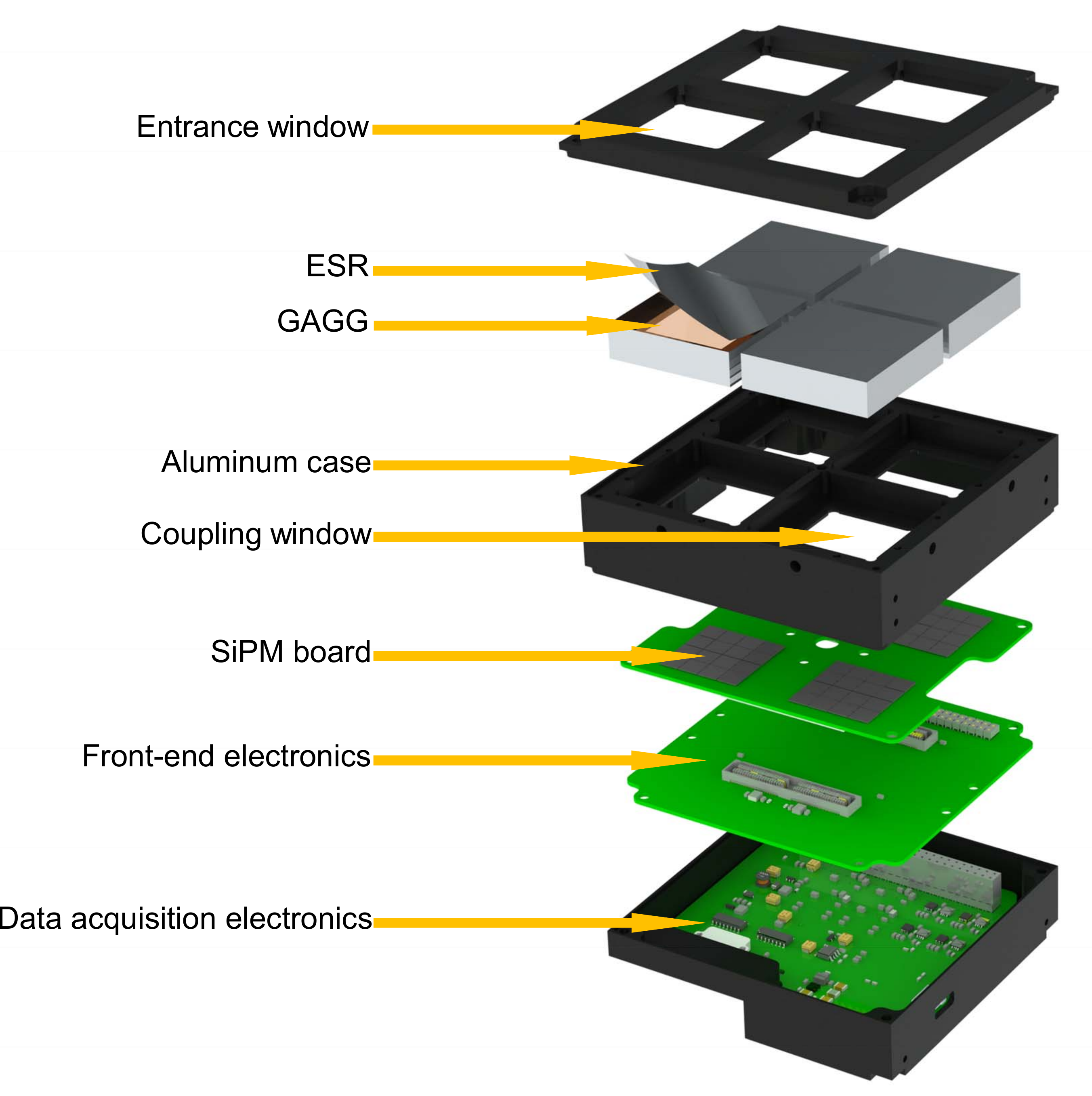}
    \caption{Exploded view of the GRID detector. Details about GRID design can be found in \cite{Wen2021}. }
    \label{fig:detector_exploded}
\end{figure}

\begin{figure}
    \centering
    \includegraphics[width=1.0\textwidth]{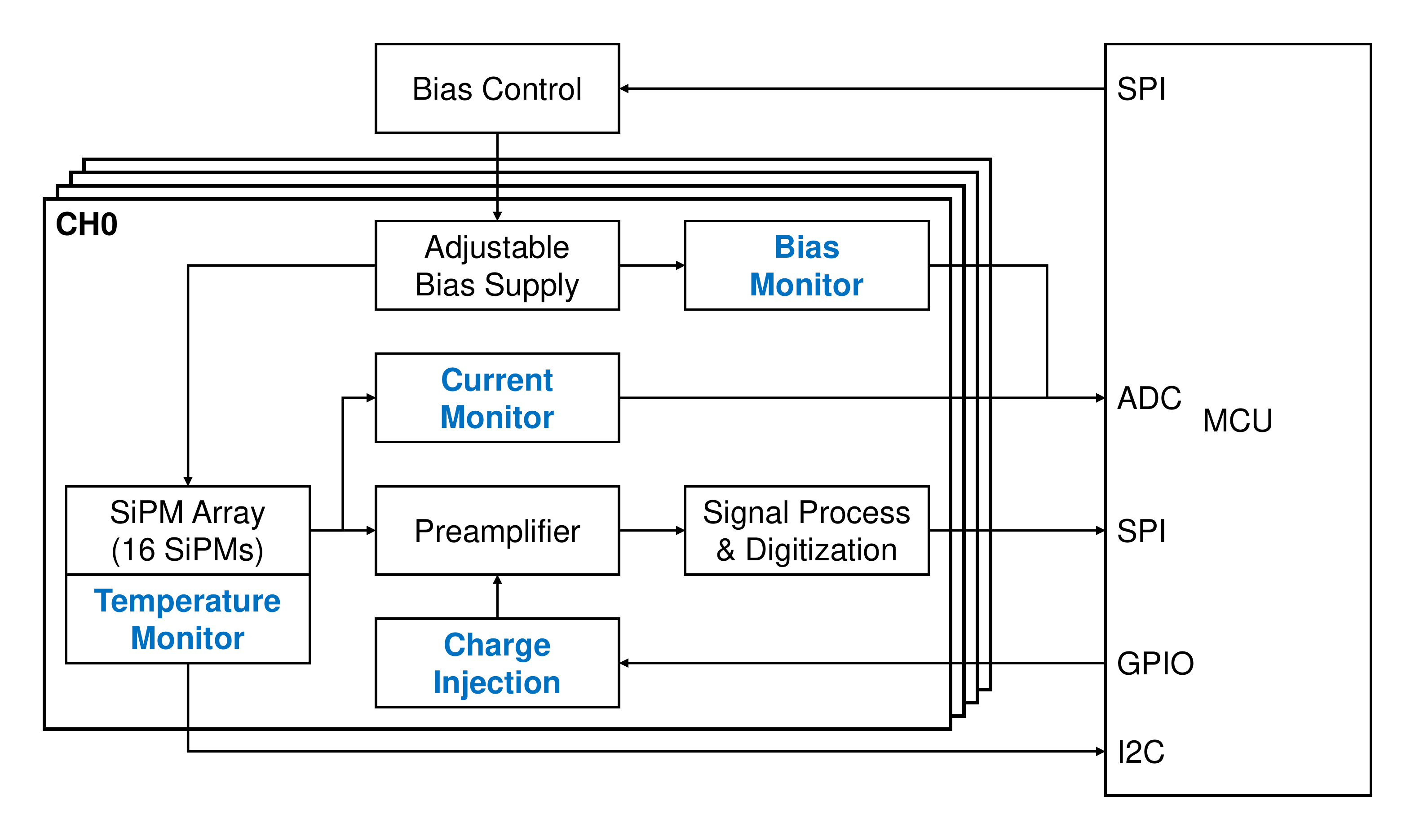}
    \caption{A simplified block diagram of the GRID detector. }
    \label{fig:block_diagram}
\end{figure}

The exploded view and simplified block diagram of the GRID detector are shown in Fig. \ref{fig:detector_exploded} and Fig. \ref{fig:block_diagram}, respectively. 
The detector comprises four independent channels designated as CH0 - CH3. 
Each channel consists of a scintillator crystal, 4 $\times$ 4 array of MicroFJ-60035-TSV\footnote{\href{https://www.onsemi.com/pdf/datasheet/microj-series-d.pdf}{MicroFJ-60035-TSV datasheet}} SiPMs, bias voltage power supply, readout electronics, and SiPM characterization circuits. 
A microcontroller unit (MCU) acts as data acquisition (DAQ) electronics to monitor and control the four channels. 

Due to limited power and payload volume on a CubeSat, there was no active cooling system in the GRID detector. 
The temperature of the GRID detector depends mainly on the overall temperature of the satellite. 
The temperature of each SiPM array was measured using a digital temperature sensor, TMP112\footnote{\href{https://www.ti.com/lit/ds/symlink/tmp112.pdf}{TMP112 datasheet}}, which offers 0.5 $^{\circ}$C accuracy. 

Moreover, a charge injection module controlled by the MCU was connected to the preamplifier, such that a constant charge pulse could be injected at a specified frequency. 
Through charge injection, with bias voltage on or off, we can analyze the noise contributions, either from the dark count noise of the SiPM, or electronic noise, in addition to other noise sources. 

\begin{figure}
    \centering
    \includegraphics[width=0.8\textwidth]{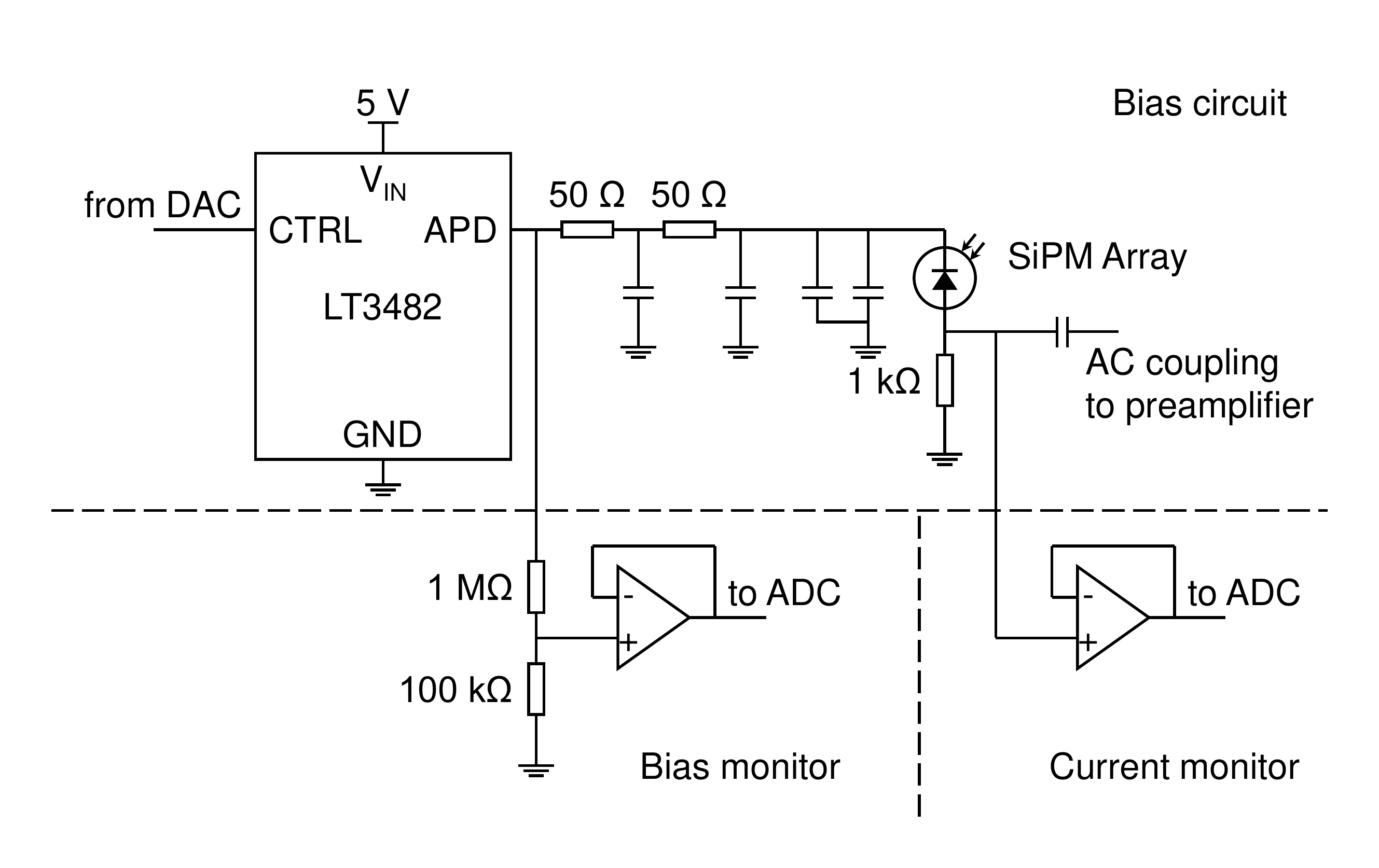}
    \caption{Schematic layout of ${I}\sim{V}$ measurement setup. }
    \label{fig:schematic_layout}
\end{figure}

Figure \ref{fig:schematic_layout} shows the schematic layout of the ${I}\sim{V}$ measurement setup. 
The 16 SiPMs in the same array were connected in parallel, which means that we can only characterize the entire array, rather than each single SiPM chip. 
The adjustable positive bias voltage on the cathode was supplied by a switching regulator, LT3482\footnote{\href{https://www.analog.com/media/en/technical-documentation/data-sheets/3482fa.pdf}{LT3482 datasheet}}, and controlled by a 12-bit digital-to-analog converter (DAC), AD5672R\footnote{\href{https://www.analog.com/media/en/technical-documentation/data-sheets/ad5672r_5676r.pdf}{AD5672R datasheet}}. 
After a resistive divider and voltage follower, it is monitored by an MCU internal 12-bit analog-to-digital converter (ADC). 
The current through the SiPMs is converted to voltage by a 1 $\rm k\Omega$ grounding resistor on the anode, then measured by the same MCU internal ADC after passing through a voltage follower. 
The difference between these two voltages is the bias voltage ($V_{\rm bias}$) of the SiPM. 
The proportional-integral-differential (PID) control method was applied to stabilize $V_{\rm bias}$. 

The current, bias voltage and temperature values were recorded every second as housekeeping data during scientific observations. 
This was used to analyze the dark current of SiPM. 
The breakdown voltage ($V_{\rm bd}$) of the SiPM was determined by ${I}\sim{V}$ measurements at different bias voltages. 
This measurement method is encapsulated in a special mode triggered by an instruction. 
Once switched to this mode, the bias voltage was successively changed to 40 preset values below and above $V_{\rm bd}$. 
The measurement lasted approximately 40 s, during which the temperature fluctuation was negligible. 
Note that when comparing the different measurements, temperature correction was performed. 

We carefully scheduled the characterization experiments and scientific observations based on the characterization setup described above. 
The payload debug phase began on 8 November 2020 after launch and finished on 19 November 2020. 
We then moved to the daily observation phase and arranged the characterization experiments once per day from 4 December 2020. 
Owing to the existence of other payloads, the scientific observation time of GRID-02 was limited to approximately 20 ks (5 - 6 h) per day. 

\section{Results}

\subsection{Breakdown voltage}
\label{sec:Vbd}

Several methods can be used to extract $V_{\rm bd}$ through ${I}\sim{V}$ measurements, as discussed in \cite{NAGY201755} and \cite{KLANNER201936}. 
For the convenience of comparison, the method described in the datasheet was applied \cite{J60035}. 
The $\sqrt{I}$ versus $V_{\rm bias}$ was fitted with a linear function above $V_{\rm bd}$, where $V_{\rm bd}$ is defined as the voltage intercept. 

\begin{figure}
    \centering
    \includegraphics[width=1.0\textwidth]{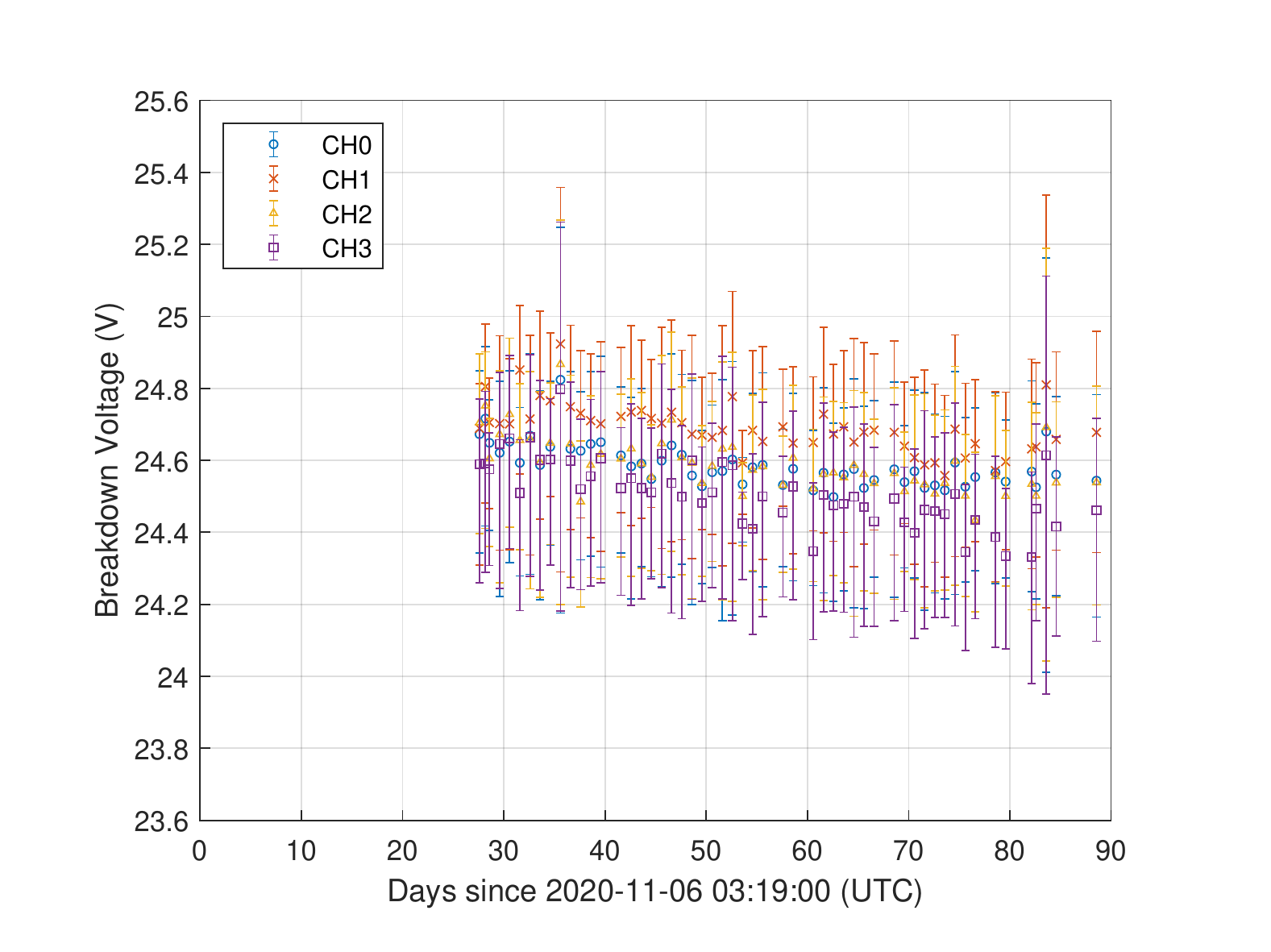}
    \caption{Breakdown voltage ($V_{\rm bd}$) of SiPM as a function of time. The values are obtained by linear fit of $\sqrt{I}$ versus bias voltage ($V_{\rm bias}$) and corrected to 5 $^{\circ}$C. The error bars show the 95\% confidence intervals. }
    \label{fig:Vbr_SensL}
\end{figure}

Temperature correction is applied through 
\begin{equation}
    \label{eq:Vbd}
    V_{\rm bd}(T_{5})=V_{\rm bd}(T)-k_{\rm bd}\cdot(T-T_{5}),
\end{equation}
where $T_{5}=5 ^{\circ}{\rm C}$ is the typical operating temperature of GRID-02 and is chosen as the reference temperature; $T$ is the absolute temperature of each ${I}\sim{V}$ measurement; $k_{\rm bd}=21.5\ {\rm mV/K}$ is the temperature coefficient of $V_{\rm bd}$ specified in the datasheet \cite{J60035}. 
This $k_{\rm bd}$ value was also used in the gain calibration procedure for GRID-02 \cite{Gao2021}. 
In Fig. \ref{fig:Vbr_SensL}, $V_{\rm bd}$ as a function of time is shown. 
Error bars represent 95\% confidence intervals. 
The systematic error due to temperature correction was relatively small; thus, it was neglected. 
No significant change in $V_{\rm bd}$ is observed. 

\subsection{Dark current}
\label{sec:Idark}

The dark current ($I_{\rm dark}$) of SiPM can be expressed by 
\begin{equation}
    \begin{aligned}
        \label{eq:Idark}
        I_{\rm dark}&=DCR\cdot(Gain\cdot{e})\cdot{ECF}\\
        &=DCR\cdot{C}_{\rm pix}\cdot(V_{\rm bias}-V_{\rm bd})\cdot{ECF},
    \end{aligned}
\end{equation}
where $DCR$ is the dark count rate, $Gain$ is the overall gain of SiPM, $e$ is the elementary charge, $ECF$ is the excess charge factor, and $C_{\rm pix}$ is the pixel capacitance including the quenching capacitance \cite{KLANNER201936}. 
The term $(V_{\rm bias}-V_{\rm bd})$ is also known as overvoltage ($V_{\rm ov}$). 

$C_{\rm pix}$ was found to be independent of temperature and radiation damage \cite{GARUTTI201969}; $ECF$, which depends on $V_{\rm ov}$, is assumed constant; $V_{\rm bias}$ is set to 28.5 V without any temperature compensation, so the temperature dependence of $V_{\rm bd}$ needs to be taken into account as discussed in Section \ref{sec:Vbd}. 
Field-Enhanced Shockley-Read-Hall (FE-SRH) model \cite{GARUTTI201969, 155882} indicates the relationship between $DCR$ and temperature as follows: 
\begin{equation}
    \label{eq:FESRH}
    DCR(T)\propto\left(1+2\sqrt{3\pi}\frac{F_{\rm eff}}{(kT)^{3/2}}{\rm e}^{\left(\frac{F_{\rm eff}}{(kT)^{3/2}}\right)^2}\right)T^2{\rm e}^{-\frac{E_{\rm a}}{kT}},
\end{equation}
where $F_{\rm eff}$ is the effective electric field strength, $k$ is Boltzmann constant, and $E_{\rm a}=0.605\ {\rm eV}$ is the activation energy. 
Because $V_{\rm bias}$ is set to 28.5 V, $F_{\rm eff}$ is assumed to remain constant. 

\begin{figure}
    \centering
    \includegraphics[width=1.0\textwidth]{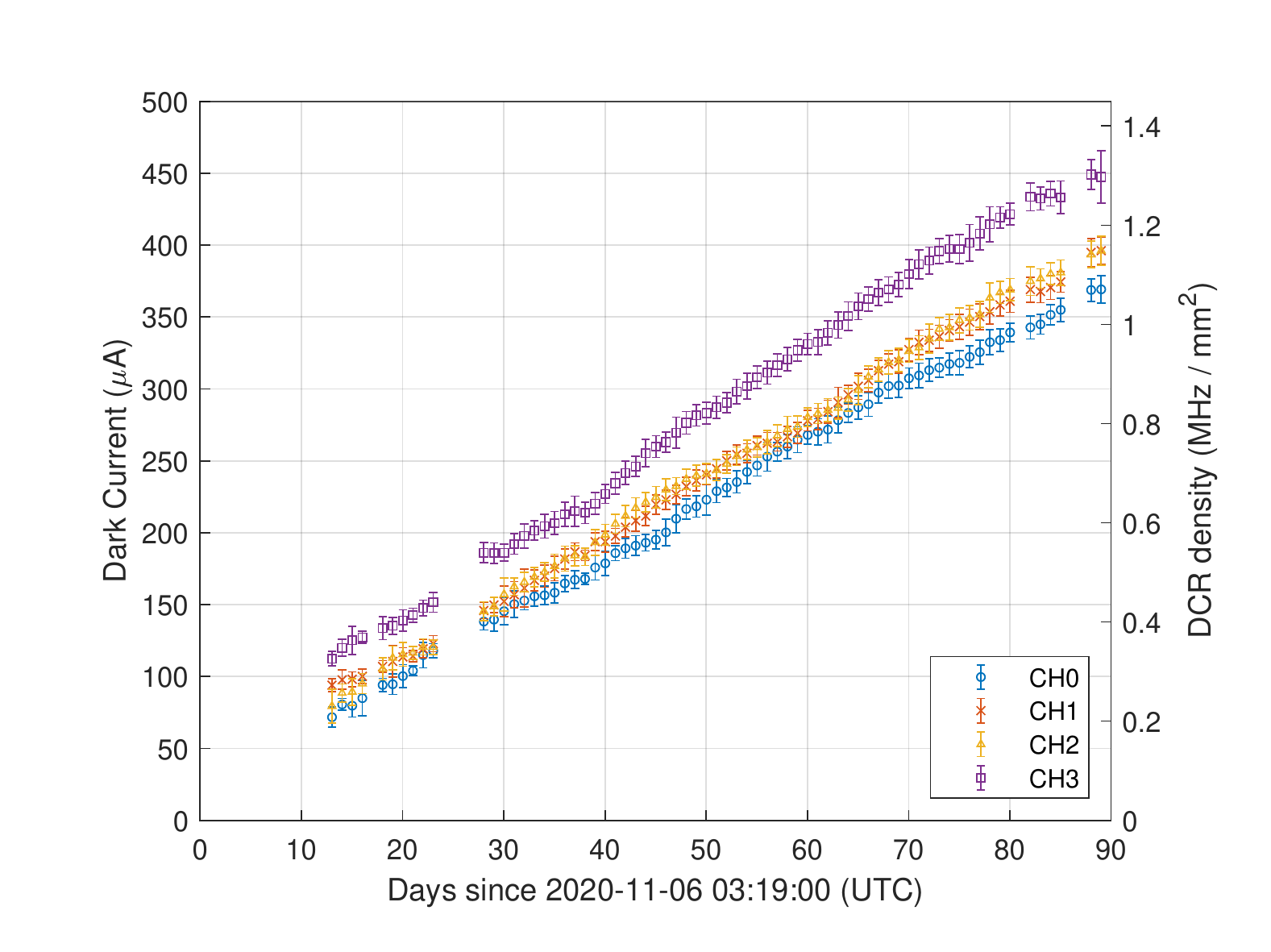}
    \caption{Dark current ($I_{\rm dark}$) at 28.5 V bias voltage as a function of time. The values are the sum of 16 SiPMs in the same channel and are unified to 5 $^{\circ}$C. }
    \label{fig:IT_p5}
\end{figure}

Finally, the $I_{\rm dark}$ expression becomes 
\begin{equation}
    \label{eq:Idark_fit}
    I_{\rm dark}(T)\propto\left(1+2\sqrt{3\pi}\frac{F_{\rm eff}}{(kT)^{3/2}}{\rm e}^{\left(\frac{F_{\rm eff}}{(kT)^{3/2}}\right)^2}\right)T^2{\rm e}^{-\frac{E_{\rm a}}{kT}}\cdot(V_{\rm bias}-V_{\rm bd}(T)).
\end{equation}
We split the data by date, and the daily change in $I_{\rm dark}$ corresponded to the average radiation damage effect per day. 
Then, we fitted the plot of $I_{\rm dark}$ versus temperature using Eq. (\ref{eq:Idark_fit}) and took $F_{\rm eff}$ as the free parameter. 
The $I_{\rm dark}$ values were unified to 5 $^{\circ}$C and are shown in Fig. \ref{fig:IT_p5} as a function of time. 

A linear fit shows that $I_{\rm dark}$ increases $\sim$93/96/98/110 $\mu$A/year per SiPM chip for CH0 - CH3 respectively. 
The increase in $I_{\rm dark}$ observed by SIRI-1 for the same SiPM operating in LEO is $\sim$132 $\mu$A/year per SiPM chip, despite that the operating temperature is unknown according to \cite{mitchell2019strontium} and \cite{MITCHELL2021164798}. 
A discussion of the dose difference between GRID-02 and SIRI-1 is provided in Section \ref{sec:Dose}. 
Furthermore, it is unclear what causes the differences among the four channels of GRID-02. 
Possible reasons include the difference between the temperatures of the SiPMs and the temperature sensor, and the mechanical damage received before or during launch. 

Assuming $ECF\approx$ 1.2 \cite{KLANNER201936} and taking into account fill factor = 0.75, total SiPM area\footnote{While in the datasheet, this is referred to as active area. See \cite{J60035}.} = 36.8 mm$^2$, $C_{\rm pix}$ = 0.186 pF \cite{J60035}, and $V_{\rm ov}$ = 3.5 V, the equivalent $DCR$ per area can be calculated from Eq. (\ref{eq:Idark}). 
The increase in $DCR$ density is $\sim$4.6 MHz/(year$\cdot$mm$^2$) at 5 $^{\circ}$C. 
We also extrapolated the fit of $I_{\rm dark}$ versus temperature to -20 $^{\circ}$C to evaluate the SiPMs working with cooling systems. 
The increase rate of $I_{\rm dark}$ at -20 $^{\circ}$C is $\sim$40 $\mu$A/year per SiPM chip, which is less than half of the value at 5 $^{\circ}$C. 
This is equivalent to $\sim$1.9 MHz/(year$\cdot$mm$^2$) increase in $DCR$ density and is in accordance with the temperature dependence of dark current expected by FE-SRH model \cite{GARUTTI201969, 155882}. 

\subsection{Dark count noise}
\label{sec:noise}

\begin{figure}
    \centering
    \subfigure[bias voltage off]{
        \includegraphics[width=0.47\textwidth]{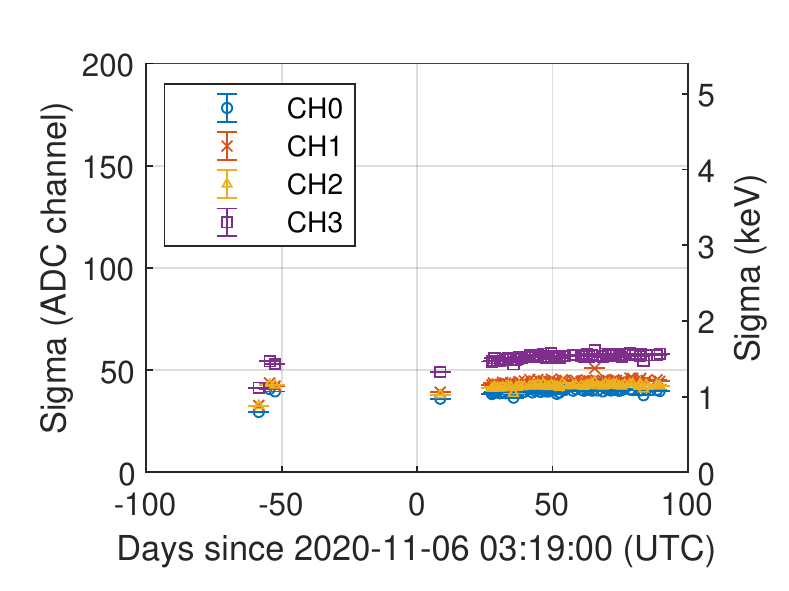}
        \label{subfig:sigma_off}
    }
    \subfigure[28.5 V bias voltage]{
        \includegraphics[width=0.47\textwidth]{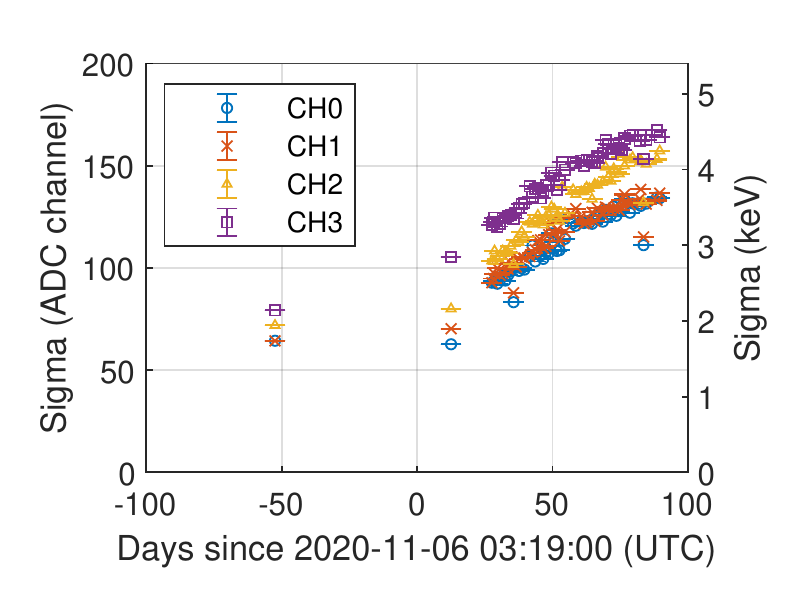}
        \label{subfig:sigma_on}
    }
    \caption{Standard deviation ($\sigma$) of the charge injection peak as a function of time. }
    \label{fig:sigma_onoff}
\end{figure}

The energy resolution and noise contributions of the GRID detector were tested via charge injection. 
The charge injection peaks with bias voltage on and off were fitted with Gaussian functions to derive the standard deviation ($\sigma$), as shown in Fig. \ref{fig:sigma_onoff}. 
Negative date values represent results obtained before launch. 
The error bars represent 95\% confidence intervals. 
The results in ADC channels are converted to equivalent energy with a scale factor $\sim$0.027 keV/channel extracted from the on-ground calibration data of GRID-02 \cite{Gao2021}. 
Note that this value is heavily dependent on the detector design, such as the electronic gain and the light yield of the scintillator crystal; thus, it is only applicable to GRID-02. 

The $\sigma_{\rm elec}$ in Fig. \ref{subfig:sigma_off} shows no significant change because it is dominated by the electronic noise and fluctuation of the injected charge itself. 
However, when a bias voltage was applied, the dark count noise became significant. 
$\sigma_{\rm total}$ gradually increases with $I_{\rm dark}$, as shown in Fig. \ref{subfig:sigma_on}. 
The total noise $\sigma_{\rm total}$ can be expressed as 
\begin{equation}
    \label{eq:sigma_total}
    \sigma_{\rm total}^2=\sigma_{\rm DC}^2+\sigma_{\rm elec}^2,
\end{equation}
where $\sigma_{\rm DC}$ is the dark count noise. 
Campbell's theorem \cite{Campbell} shows its relationship with dark count rate ($DCR$) and $I_{\rm dark}$: 
\begin{equation}
    \begin{aligned}
        \label{eq:Campbell}
        \sigma_{\rm DC}^2&=DCR\cdot(Gain\cdot{e})^2\cdot\int{h}^2(t){\rm d}t\\
        &\propto{I}_{\rm dark}\cdot(V_{\rm bias}-V_{\rm bd}),
    \end{aligned}
\end{equation}
where $h(t)$ is the unit impulse response of the readout electronics. 
Note that $I_{\rm dark}$ here is the actual measurement result and should not be temperature-corrected. 
A discussion on how to evaluate the usable lifetime of the instrument based on these results is provided in Section \ref{sec:Lifetime}. 

\section{Discussions}

\subsection{Dark current estimation}

Several previous studies have found a linear relationship between $I_{\rm dark}$ and radiation damage measured by dose or particle fluence \cite{LI201663, MITCHELL2021164798, GARUTTI201969}. 
By assuming that the radiation damage of SiPMs in the GRID detector is linearly correlated with time, and the $F_{\rm eff}$ in Eq. (\ref{eq:Idark_fit}) is independent of radiation damage, Eq. (\ref{eq:Idark_fit}) is modified as 
\begin{equation}
    \begin{aligned}
        \label{eq:Idark_fit_time}
        I_{\rm dark}(Time,T)&=(a_{\rm dc}\cdot{Time}+b_{\rm dc})\\
        &\mathrel{\phantom{=}}\cdot\left(1+2\sqrt{3\pi}\frac{F_{\rm eff}}{(kT)^{3/2}}{\rm e}^{\left(\frac{F_{\rm eff}}{(kT)^{3/2}}\right)^2}\right)T^2{\rm e}^{-\frac{E_{\rm a}}{kT}}\\
        &\mathrel{\phantom{=}}\cdot\left(V_{\rm bias}-V_{\rm bd}(T)\right),
    \end{aligned}
\end{equation}
where $a_{\rm dc}$, $b_{\rm dc}$, and $F_{\rm eff}$ are free parameters and $Time$ is the elapsed time after launch in days. 
A joint fit of $I_{\rm dark}$ versus time and temperature provides an estimate of $I_{\rm dark}$ at specified conditions. 
The fitting results are presented in Table \ref{tab:Idark_time_temp}. 

\begin{table}
    \begin{center}
        \caption{Fitting results of Eq. (\ref{eq:Idark_fit_time}). }
        \begin{tabular}{cccc}
            \hline
                & $a_{\rm dc}$  & $b_{\rm dc}$  & $F_{\rm eff}/k^{3/2}$ \\
            \hline
            CH0 & 738.184       & 4227.74       & 10124.9   \\
            CH1 & 642.893       & 5072.58       & 10286.3   \\
            CH2 & 733.990       & 5295.85       & 10178.1   \\
            CH3 & 871.504       & 8740.97       & 10122.3   \\
            \hline
        \end{tabular}
        \label{tab:Idark_time_temp}
    \end{center}
\end{table}

By substituting these parameters into Eq. (\ref{eq:Idark_fit_time}), we obtain an approximate empirical equation around room temperature which describes the $I_{\rm dark}$ of GRID-02 at 28.5 V bias voltage, and at a specified time and temperature: 
\begin{equation}
    \label{eq:Idark_Time_T_pred}
    I_{\rm dark}(Time,T)=16\cdot(0.2678\cdot{}Time+2.091)\cdot{}{\rm e}^{0.03475\cdot(T-T_{5})},
\end{equation}
where $T_{5}=5 ^{\circ}{\rm C}$ is the reference temperature of GRID-02. 
The unit of $I_{\rm dark}$ is $\mu$A, and the first term (16) means the result is the sum of the 16 SiPMs, as in the GRID detector. 
The exponential term shows that $I_{\rm dark}$ doubles for every $\sim$20 $^{\circ}$C temperature increase. 

\subsection{Preliminary dose estimation}
\label{sec:Dose}

\begin{figure}
    \centering
    \includegraphics[width=1.0\textwidth]{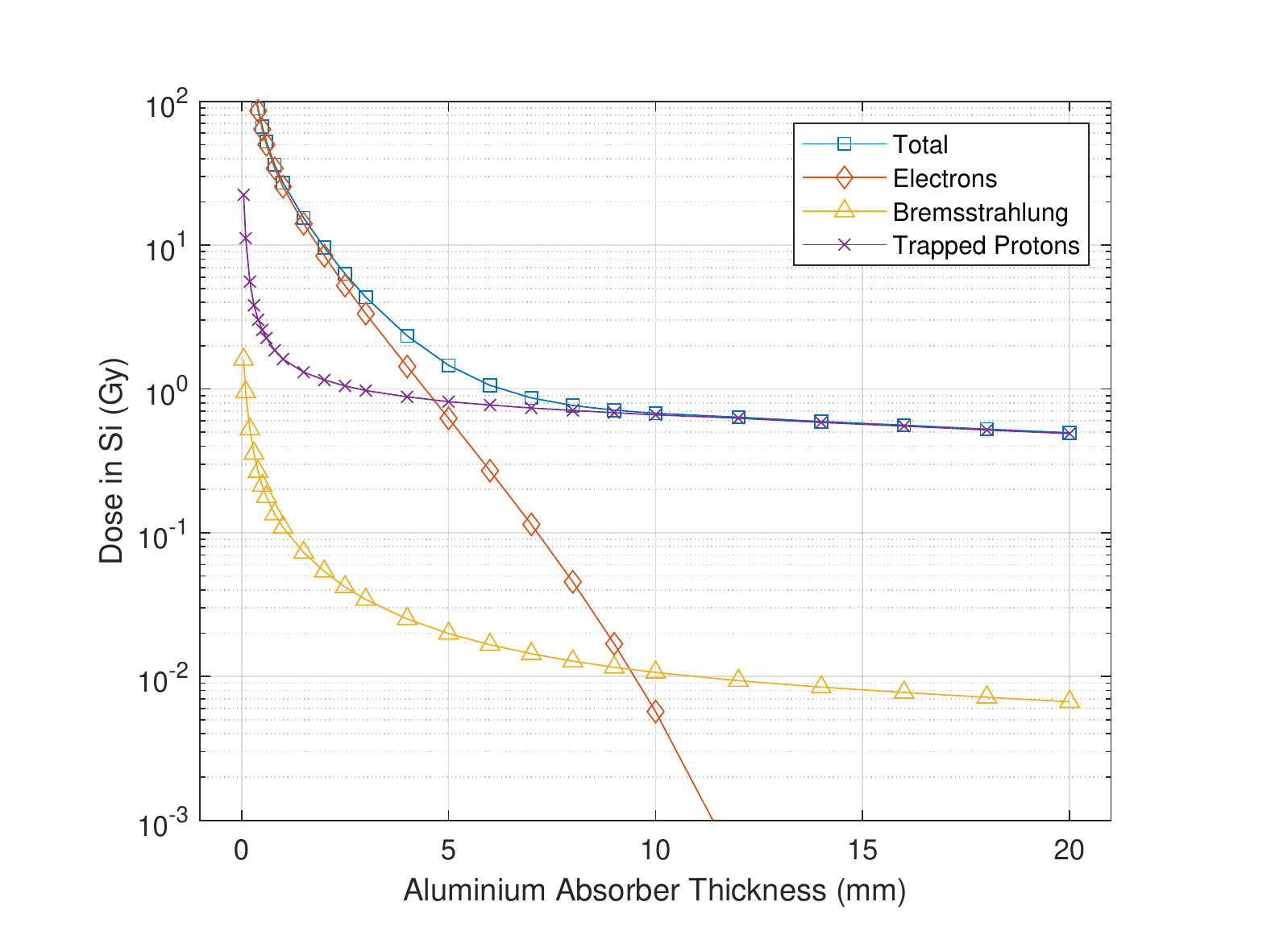}
    \caption{Annual dose in silicon as a function of aluminum shielding thickness for GRID-02 given by SHIELDOSE-2 from SPENVIS \cite{SPENVIS, SHIELDOSE2}. Trapped radiation models for protons and electrons are AP-8 and AE-8 (solar minimum), respectively. }
    \label{fig:dose}
\end{figure}

The radiation damage effect on SiPMs depends on the irradiation particle type, energy, and fluence, and is traditionally evaluated by the dose or 1 MeV neutron equivalent fluence \cite{GARUTTI201969}. 
Referring to the method used by SIRI-1 \cite{MITCHELL2021164798}, we provided a preliminary estimate of the dose. 
The SPENVIS online program \cite{SPENVIS} can generate particle fluxes with orbital information, as listed in Table \ref{tab:missions}. 
AP-8 and AE-8 (solar minimum) trapped particle radiation models were selected for protons and electrons, respectively. 
Then, using the SHIELDOSE-2 model \cite{SHIELDOSE2}, SPENVIS calculates the cumulative dose for a given period (1 year in this case) at different shielding thicknesses, as shown in Fig. \ref{fig:dose}. 
Considering that the equivalent shielding thicknesses in all directions are no less than 2 cm for GRID-02, the asymptotic value of the total dose, $\sim$0.5 Gy, was used as an approximation of the annual dose. 
The empirical equation for $I_{\rm dark}$ becomes: 
\begin{equation}
    \label{eq:Idark_Dose_T_pred}
    I_{\rm dark}(Dose,T)=16\cdot(195.5\cdot{}Dose+2.091)\cdot{}{\rm e}^{0.03475\cdot(T-T_{5})},
\end{equation}
where $Dose$ is the cumulative dose (Gy). 
Using the method described in Section \ref{sec:Idark}, the equivalent $DCR$ density can be given as 
\begin{equation}
    \label{eq:DCR_Dose_T_pred}
    DCR(Dose,T)=(9.101\cdot{}Dose+0.09733)\cdot{}{\rm e}^{0.04019\cdot(T-T_{5})},
\end{equation}
where the unit of $DCR$ is MHz/mm$^2$. 
We caution that these equations are applicable only when the bias voltage is 28.5 V. 
Furthermore, a detailed Monte Carlo dose simulation could be performed with the mass model of the satellite and in-orbit radiation environment models. 

To test this empirical equation, the dark current increase rate calculated by Eq. (\ref{eq:Idark_Dose_T_pred}) was compared with the measurement results of SIRI-1 and GRID-04. 
These values were in good agreement with each other, as listed in Table \ref{tab:Idark_pred}. 
This estimate of the dark current can then be used to calculate the noise increase and evaluate the usable lifetime of the instrument. 

\begin{table}
    \begin{center}
        \setlength{\leftskip}{-54pt}
        \caption{Measured and estimated dark current increase rate of SIRI-1 and GRID-04. }
        \begin{tabular}{ccccccc}
            \hline
                        & Operating         & Operating     &                                       & Dose in   & \multicolumn{2}{c}{Dark current increase rate}    \\
            Mission     & temperature       & voltage       & Orbit                                 & silicon   & \multicolumn{2}{c}{per SiPM chip ($\mu$A/year)}   \\
            \cline{6-7}
                        & ($^{\circ}$C)     & (V)           &                                       & (Gy)     & Measured              & Estimated                  \\
            \hline
            SIRI-1      & 7.75              & 28.5          & 567 $\times$ 589 km 97.7$^{\circ}$    & 0.9       & 132                   & 194                       \\
            GRID-04     & 5                 & 28.5          & 523 $\times$ 550 km 97.5$^{\circ}$    & 0.9       & 182                   & 176                       \\
            \hline
        \end{tabular}
        \label{tab:Idark_pred}
    \end{center}
\end{table}

\subsection{Usable lifetime evaluation of SiPM}
\label{sec:Lifetime}

For a detector system readout using the SiPM, the dark count noise is one of the major contributors to the overall noise level ($\sigma_{\rm total}$). 
As the dark current increased with radiation damage, the noise level also increased. 
In addition, the energy resolution and low energy threshold significantly deteriorate simultaneously, both of which are critical performance parameters for usable lifetime evaluation of this type of instrument. 
On the one hand, for GRID-02 which is in orbit, by measuring the degradation of the energy resolution, the measurement results of the dark current and the law of its increase can be confirmed. 
On the other hand, for any detector system in advance of launch, the experimentally obtained relationship between the energy resolution and dark current can be used to evaluate its usable lifetime in space. 
More specifically, the energy resolution (i.e., the noise level in $\sigma$) can be measured for different dark currents on the ground. 
Then, with the estimate of dose and the dark current given by the empirical equation above, the increase rate of the noise level can be obtained, and the usable lifetime of the instrument can be predicted. 

\begin{figure}
    \centering
    \includegraphics[width=1.0\textwidth]{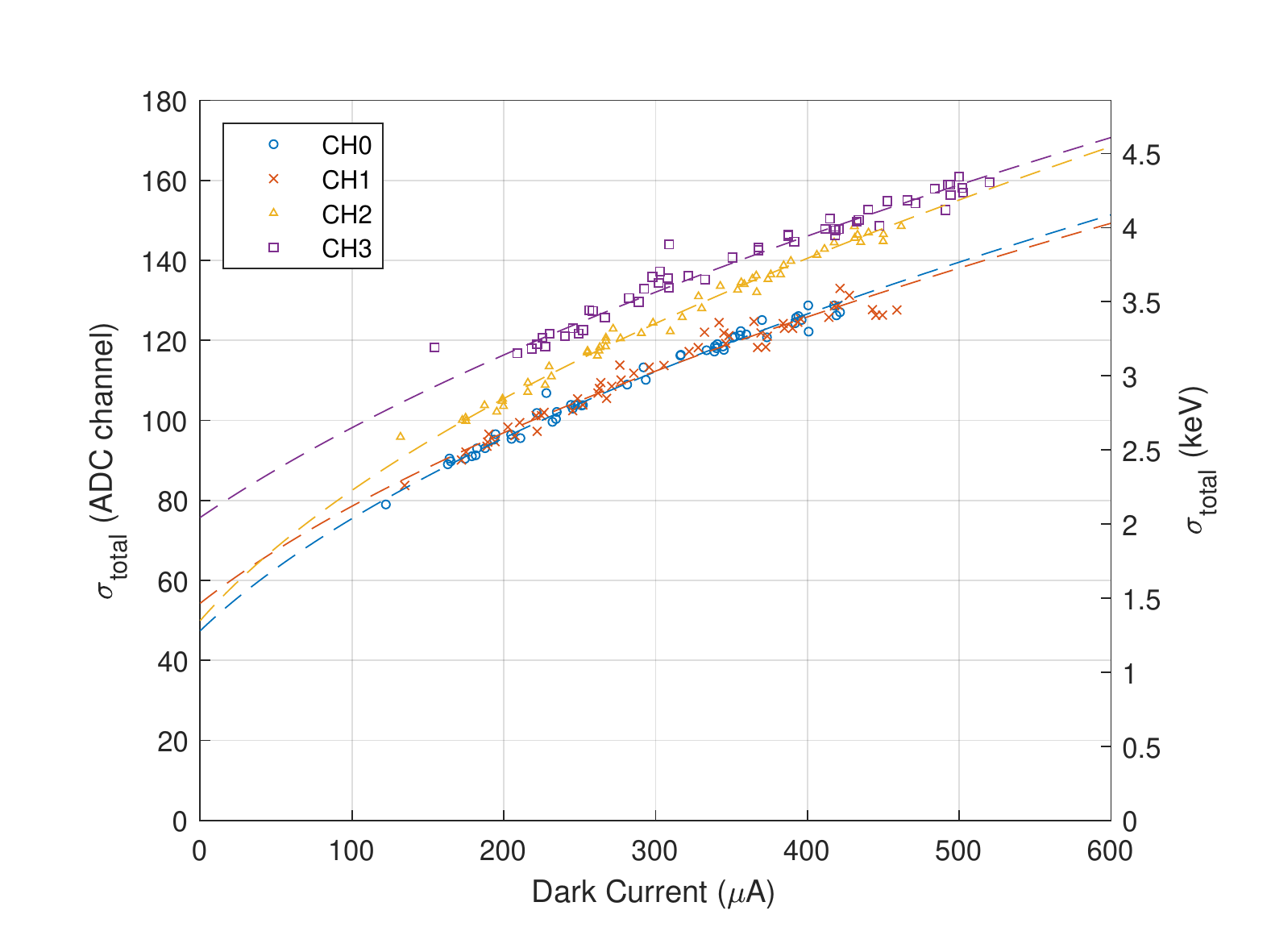}
    \caption{Detector total noise ($\sigma_{\rm total}$) versus dark current ($I_{\rm dark}$). The current are origin values without any correction. }
    \label{fig:sigmatotal_I}
\end{figure}

Here, we use GRID as an example to demonstrate this procedure. 
In Fig. \ref{fig:sigmatotal_I}, we show a plot of $\sigma_{\rm total}$ versus $I_{\rm dark}$ and fit them with 
\begin{equation}
    \label{eq:sigma_Idark_fit}
    \sigma_{\rm total}=\sqrt{a_{\rm n}\cdot{}I_{\rm dark}\cdot(V_{\rm bias}-V_{\rm bd})+b_{\rm n}},
\end{equation}
which is derived from Eqs. (\ref{eq:sigma_total}) and (\ref{eq:Campbell}). 
The free parameters, $a_{\rm n}$ and $b_{\rm n}$, are heavily dependent on the readout electronics design, such as the readout time constant and the gain. 
The results indicated an average noise ($\sigma_{\rm total}$) increase of $\sim$4.7 keV/mA for GRID-02 over the first months. 
Considering the roughly 1.6 mA/year increase in dark current, the noise increase rate over time is $\sim$7.5 keV/year. 
The energy resolution (FWHM = 2.355$\cdot\sigma$) and low energy threshold (6$\sigma$ for GRID-02) could be calculated as well. 
Once they no longer meet the performance requirements of scientific goals, the instrument is declared to have reached the end of its life. 
Note that similar results could be obtained with on-ground calibrations, for example, by measuring the energy resolution or performing charge injection at different temperatures (and thus different dark currents), so that this noise analysis and usable lifetime evaluation procedure could be carried out before launch, when there is still a chance to improve the design. 

\subsection{Methods of improve signal-to-noise ratio}

As discussed in Section \ref{sec:noise}, the noise of the SiPM owing to radiation damage is dominated by dark count noise, as described by Eq. (\ref{eq:Campbell}). 
Here, we discuss qualitative methods for improving the signal-to-noise ratio (SNR). 

The first term is $DCR$ which depends on temperature, as shown in Eq. (\ref{eq:FESRH}). 
This leads to a straightforward method, which is to lower the operating temperature. 
This might be difficult for CubeSats but is applicable to larger satellites. 
The exponential term in Eq. (\ref{eq:DCR_Dose_T_pred}) provides a practical approximation around room temperature that $DCR$ is reduced by half for every $\sim$17 $^{\circ}$C decrease in temperature. 

The second term is $Gain$, which is proportional to $V_{\rm ov}=V_{\rm bias}-V_{\rm bd}$. 
As a critical operating parameter, the domino effect is significant when $V_{\rm ov}$ decreases. 
On the one hand, noise reduces with a decrease in the $Gain$ and $DCR$. 
On the other hand, the signal also decreases owing to a decrease in the $Gain$ and photon detection efficiency ($PDE$). 
Care must be taken to determine the optimum value of $V_{\rm ov}$. 

The third term represents the response of the readout electronics. 
In general, a shorter readout time constant leads to lower dark count noise. 
However, an overly short readout time constant might cause problems such as a larger ballistic deficit and more electronic noise. 
The lower limit is usually due to the scintillation decay time. 

\section{Conclusion}

With the successful observation of several gamma-ray bursts using GRID-02, the feasibility of SiPM in space applications was once again proven. 
However, significant radiation damage indicates that a usable lifetime evaluation requires more attention from the very beginning of the instrument design stage. 
As a CubeSat detector equipped with scintillator crystals and SiPMs optimized for gamma-ray detection, GRID was specifically designed with SiPM characterization setups to conduct in-orbit characterization experiments. 
The characterization results of GRID-02 for the first few months show an increase in dark current of $\sim$100 $\mu$A/year per SiPM chip (model MicroFJ-60035-TSV) at 28.5 V and 5 $^{\circ}$C. 
This is equivalent to $\sim$4.6 MHz/(year$\cdot$mm$^2$) increase in $DCR$ density and leads to the increase in the overall noise level (sigma) of $\sim$7.5 keV/year. 
Based on the SPENVIS online program, the estimated dose of GRID-02 is $\sim$0.5 Gy/year. 
An approximate empirical equation around room temperature is given for the dark current estimation at a specified dose and temperature. 
The measured dark currents of SIRI-1 and GRID-04 were used to test this equation, and showed good agreement. 
This equation also implies that the reduction in operating temperature could significantly mitigate the increase in dark current and noise, which means that a cooling system is highly recommended. 
Carefully designed readout electronics and optimum operating voltage also improve the signal-to-noise ratio. 

\section*{Acknowledgement}

This work is supported by the Tsinghua University Initiative Scientific Research Program and the National Natural Science Foundation of China (Grant No. 11961141015). 
The authors would like to thank Dr. Shaolin Xiong and his colleagues of the GECAM satellite group from the IHEP of CAS, for open and fruitful discussions on topics related to this work. 

\bibliography{References}

\end{document}